\documentstyle[epsfig,aps,preprint,prbbib]{revtex}
\def \E{\langle {\bf E} \rangle}
\def \e0{\langle {\bf E}^0 \rangle}

\def \dei{ {\delta {\bf i}}}
\def \di{\langle {\delta {\bf i}} \rangle}
\def \bE{{\bf E}}
\def \bi{{\bf i}}
\def \bA{{\bf A}}

\def \br{{\bf r}}
\def \cc{{\cal C}}
\def \vtot{{V_{tot}}}
\def \vin{{V_{p}}}
\def \vout{{V_{sol}}}
\def \sp{{S_p}}
\def \rhof{{\rho_f}}
\def \se{{S_{e}}}
\def \bnabla{\mbox{\boldmath $\nabla$}}

\begin{document}

\title{The Static Dielectric Constant of a Colloidal Suspension.}

\author{C.-Y. D. Lu\\
Departments of Physics and Chemistry, Center of Complex System, \\ National Central University, Chung-li 320,  Taiwan \\E-mail: dlu@phy.ncu.edu.tw
}
\maketitle

\begin{abstract}

We derive an expression for the static dielectric constant of the colloidal suspensions based on the electrokinetic equations. The analysis assumes that the ions have the same diffusivity, and that the double layer is much thinner than the radius of curvature of the particles. It is shown that the dielectric increment of the double layer polarization mechanism is originated from the free energy stored in the salt concentration  inhomogeniety. We also show that the dominant polarization charges in the theory are at the electrodes, rather than close to the particles.

\end{abstract}



\pagebreak

\pagestyle{plain}

\section{Introduction}

Since its invention \cite{Dukhin,Fixman,De Lacey}, the double layer polarization theory of charged colloidal suspension is well known for its complexity. 
In the theory, the dielectric function is calculated by analysing the out of phase (charging) electric current at finite frequencies. This approach makes the theory hard to grasp.  Although equivalent, the out of phase current is not as close to the intuitive thinking as the dipole built up under the applied field. In particular the location of the polarization charge is not obvious from the theory. Some researchers even wrongly assumed that the dominant polarization charge is around the particle when trying to simplify the theory.

Another consequence of using current to do calculation is that the static dielectric constant is obtained indirectly, via taking the zero frequency limit of the dynamic ac calculation. Normally one would like to be able to calculate the static dielectric constant statically, because such calculation will be easier to understand than the dynamic one.  However, for many years there is no static calculation which reproduces the known result.

In a recent work  \cite{Grosse97}, Grosse and Shilov, based on the equilibrium thermodynamic argument, suggested that the zero frequency dielectric constant should be obtained by calculating the free energy stored, as the form of the non-uniform salt concentration around the particles. 
For the system of $z:z$ electrolyte, their suggestion reads
\begin{equation}
\vtot \frac{\Delta \epsilon }{2}  E^2= \Delta F
=\frac{kT}{2}\int \left[ \frac{(\delta c^+)^2}{ C}+\frac{(\delta c^-)^2}{ C} \right] dV
\label{Shilov}
\end{equation}
where $\vtot$ is the total volume,  $E$ is the applied electric field strength. The static dielectric increment $\Delta \epsilon $ is the  static absolute dielectric constant subtracts the contribution from the simple electrolyte solution.  
The free energy stored  $\Delta F$, is calculated by integrating the free energy changes due to the counterion and coion concentration perturbations $\delta c^+$, $\delta c^-$  outside the double layer, where $ C$ is the equilibrium ion concentration in the bulk, $k$, $T$ are the Boltzmann constant and the temperature respectively. In fact, the ions concentrations are the same outside the double layer, therefore the free energy is stored in  the form of non-uniform salt concentration distribution. 

Grosse and Shilov have checked that eq.(\ref{Shilov}) is obeyed for the well known solution of the dilute charged sphere suspension. They further used eq.(\ref{Shilov}) to calculate the static dielectric increment of dilute ellipsoids, which is not easy if the original calculation method is used. 

In this work, we prove eq.(\ref{Shilov}) (with some corrections from  the double layer region) based on the electrokinetic equations, so that its general validity is established on the same basis as the double layer polarization theory. We consider only the simple case where the two ions have the same mobility, but put  no restriction on the $\zeta$ potential value. By analysing the  charging current, we also show explicitly that the polarization charges are close to the two electrodes. Below, we summarize the thin double layer theory in section II. The charging electric current density and the dipole are analysed in  section III. Eq.(\ref{Shilov}) is derived in section IV. A short summary concludes the paper.

\section{The Thin Double Layer Theory}

The standard electrokinetic theory \cite{Dukhin,Fixman,De Lacey,Hunter}  describes the ion dynamics ($\delta c^\pm$) by the Smoluchowski equations, the electrostatic potential ($\phi$)  follows the Poisson's equation, and the fluid velocity obeys the Navier-Stokes equation with an electric body force given by the product of the local charge density and the electric field.

The analysis of the dynamic equation proceeds differently within and outside the double layer (the bulk), where different approximations been appropriate \cite{Dukhin,Fixman,O'Brien83}. The bulk equations to be used later are summarized in subsection \ref{bulk} below (which apply for the $z:z$ electrolyte with the equal ion mobility). The partial solutions from the two regions are matched smoothly at the boundary of the double layer to give the full solution.

In the case of a thin double layer, where the radius of curvature of the particle is much larger than the double layer thickness, the double layer region can be approximated as a flat double layer. 
The solution within the flat double layer is known and the matching conditions act effectively as the boundary conditions for the quantities outside the double layer. In subsection \ref{bc} below we summaries the boundary conditions to be used.

\subsection{The Bulk Equations}
\label{bulk}
It is well known that, in the case of $z:z$ electrolyte with the equal ion mobility, the solution becomes charge neutral a few Debye lengths away from any interface \cite{Lifshitz}. The salt perturbation does not carry charge, therefore  uncouples with the potential. The unperturbed salt has a uniform distribution so the flow does not couple with the salt perturbation by convection (to the first order perturbation). The bulk solution is charge neutral to begin with, so the potential does not produce force to couple with the flow. In the bulk  one only needs to solve the Laplace equation
\begin{equation}
\nabla^2 \phi=0
\label{phi}
\end{equation}
for the potential, and
 the simple diffusion equation
\begin{equation}
\partial_t \delta c=D \nabla^2 \delta c
\label{c}
\end{equation}
for the salt perturbation  $\delta c \equiv (\delta c^++ \delta c^-)/2$. The diffusivity $D$ is the same for both ions. Since the flow velocity couples passively to other quantities (via the boundary condition at the double layer), we do not need to calculate it for analysing the dielectric response.

The conduction current density $\bi$ in the bulk is local and Ohmic, obeying
\begin{equation}
\bi=K \bE
\label{Ohmic}
\end{equation}
The conductivity is given by $K=2D q^2  C/kT$, where $q$ is the counterion charge. Note that eq.(\ref{Ohmic}) is justified for the electrolyte with the equal ion mobilities, since the perturbed salt is charge neutral. For systems with different ion mobilities, there is also a (small) current density controlled by the salt, which is not proportional to the local electric field. 
The fact that ions in general have different mobilities complicate the analysis very much (all three eqs.(\ref{phi})(\ref{c})(\ref{Ohmic}) will be modified). The correction in the dielectric response \cite{Hinch}, however, seems perturbative. We will therefore focus on the simplest system in this work.


\subsection{The Double Layer Boundary Conditions}
\label{bc}

Define the perturbed surface charge  density as $\sigma=q\int_0^{\alpha/ \kappa} ( \delta c^+-\delta c^-) dr_n$ where $r_n$ is the coordinate in the direction normal to the interface (which is $r_n=0$).  
By analysing the ion conservations within the thin double layer, and assuming that no ion fluxes penetrate  into the particles, one obtains the conservation equation for the surface charge density 
\begin{equation}
\partial_t \sigma =K \nabla_n  \phi +Kk_{11}\nabla^2_\perp  \phi+Kk_{12}\nabla^2_\perp \cc
\label{sigma}
\end{equation}
where $\cc \equiv kT \delta c/q { C}$. The Debye length is $\kappa^{-1}=(2 q^2  C / \epsilon kT)^{1/2}$ where $\epsilon$ is the dielectric constant of the solvent. Note that $\sigma$ is not sensitive to the exact value of the cut off $\alpha$, whose plausible value is between $1$ to $10$.
The surface conductances are expressed as the bulk conductance $K$ times the relative conductances $k_{ij}$s, to make some formulae neater.

Using a similar analysis, the perturbed surface excess ion density $\gamma=\int_0^{\alpha/ \kappa} [ \delta c^++ \delta c^--2 \delta c(r_n=\alpha / \kappa )] dr_n$ obeys
\begin{equation}
q\partial_t \gamma =K \nabla_n  \cc +Kk_{21}\nabla^2_\perp  \phi+Kk_{22}\nabla^2_\perp \cc
\label{gamma}
\end{equation}
Here $\gamma$ is also not sensitive to the exact value of the cut off.

The surface conductances $k_{ij}$ have been calculated \cite{Dukhin,Fixman}, which include the effect of the convection current inside the double layer. There is also an Onsager relation $k_{12}=k_{21}$ as can be checked in the microscopic expressions \cite{Fixman}. There are also other boundary conditions for the flow field, and constitutive equations for $\sigma$, and $\gamma$. However we will only need eqs.(\ref{sigma})(\ref{gamma}) in this work.

Note that in the double layer polarization theory, the slow polarization of the double layer is controlled by the salt diffusion. Therefore the time derivative terms $\partial_t \sigma$, $\partial_t \gamma$ in eqs.(\ref{sigma})(\ref{gamma}) are often small, hence neglected. In this paper the time derivatives are kept, so that the interface polarization \cite{O'Brien86} (Maxwell-Wagner theory) is included.

\section{Charging Current density and the Dipole}
In this section, we decompose the total current density into the charging current density, which stores free energy, and the conducting current density, which is dissipative. We also show explicitly that the large dipole comes from the polarization charge close to the electrodes, rather than from the charge close to  the particles.

\subsection{Averaged current density}

Before we decompose the total current density, we first express the averaged current density as did by  De Lacey and White \cite{De Lacey}. The final expression has the advantage that it is valid independent of the specific current density constitutive relation close to or within the charge particles.

\begin{eqnarray}
\vtot \langle i_a \rangle &=&\int_\vout i_a dV +\int_\vin i_a dV \nonumber \\
&=&\int_\vout i_a dV +\int_\vin \nabla_b (r_a i_b) dV -\int_\vin r_a {\bf \nabla} \cdot \bi \ dV \nonumber \\
&=& \left( \int_\vout K E_a dV  \right) +\int_\sp r_a \bi \cdot d \bA +\int_\vin r_a \partial_t \rhof dV \nonumber \\
&=& \left( K\int_\vtot  E_a dV - K\int_\vin E_a dV \right) +\int_\sp r_a \bi \cdot d \bA +\int_\vin r_a \partial_t \rhof dV \nonumber \\
&=& K\int_\vtot  E_a dV +K\int_\sp \left( \phi \delta_{ab}-r_a \nabla_b \phi \right) dA_b
+\int_\vin r_a \partial_t \rhof dV
\label{bi} 
\end{eqnarray}
We define a two-components surface $\se$, which is inside the solution and very close to the two electrodes (see Fig. 1). The surface $\se$ encloses the total space $\vtot$, which is further partitioned by the surface $\sp$ into $\vtot=\vin+\vout$.    $\sp$ is a (multi-components) mathematical closed surface which includes the charged particles and their double layer (Fig.~1).  
The space $\vin$ is enclosed by the surface $\sp$.  The volume $\vout$ is the space between the two surfaces $\sp$ and $\se$. The differential area vector $d \bA$ is pointing into $\vout$ for both surfaces $\se$ and $\sp$.

In the above derivations, the space integral is decomposed into two term. The integral on $\vin$ is transformed into two term by the identity $i_a=\nabla_b(r_ai_b)-r_a \nabla_b i_b$. To get the third line we have assumed that the current in the volume $\vout$ obeys the Ohm's law $\bi=K \bE$. The second term is obtained by the Gauss' theorem. In deriving the last term,  we have not assumed any constitutive relation for the current density in $\vin$. The only property used inside $\vin$ is the charge conservation $\partial_t \rhof =- \nabla \cdot \bi$ where $\rhof$ is the free charge density. In the fourth line the integral on $\vout$ is written as $\vtot-\vin$. In the last line, the relation $E_a=-\nabla_a \phi$ and the Gauss' theorem are used. We also assumed that the Ohm's law applies on the surface $\sp$.

\subsection{Current density decomposition}

For each instance, we define the conducting current density $\bi^0$ as the current density of a reference  system, which is at a steady state where the averaged electric field  is the same as the original system at that time. The charging current density $\delta \bi$ is the difference between the total current density and the conducting current density. Given a time evolution of the applied field, for example an adiabatic charging process,  the conduction current density as defined  does not change sign under the time reversal operation. Whereas the charging current density changes sign under the time reversal.

The reference system has the fields $\phi^0$, $\cc^0$ which both obey the Laplace equations,
and the steady state boundary conditions
\begin{eqnarray}
K \nabla_n \phi^0 +Kk_{11}\nabla^2_\perp  \phi^0+Kk_{12}\nabla^2_\perp \cc^0 =0
\label{sigma0}
\\
K \nabla_n  \cc^0 +Kk_{21}\nabla^2_\perp  \phi^0+Kk_{22}\nabla^2_\perp \cc^0 =0
\label{gamma0}
\end{eqnarray}

The analysis similar to (\ref{bi}) on the conducting current $\bi^0$ gives
\begin{eqnarray}
\vtot \langle \bi^0 \rangle = K\int_\vtot  \bE^0 dV +K\int_\sp \left( \phi^0 {\bf 1}-\br {\boldmath \nabla} \phi^0 \right) \cdot d \bA
\label{i0}
\end{eqnarray}
where ${\bf 1}$ is the identity matrix. Equation (\ref{i0}) has one term less than eq.(\ref{bi}) because  $\nabla \cdot \bi^0=0$.

Define the charging current density as  $\delta \bi=\bi-\bi^0$.
Noticing that $\int_\vtot \bE dV =\int_\vtot  \bE^0 dV $ (by the definition of $\phi^0$), we subtract eq.(\ref{i0}) from eq.(\ref{bi}) to cancel their first terms to give
\begin{eqnarray}
\vtot \di =K \int_{\sp} \left( \delta \phi {\bf 1}-\br {\bf \nabla} \delta \phi \right) \cdot d {\bf A}+  \int_{\vin} \br \partial_t \rhof dV
\label{bdi}
\end{eqnarray}
where $\delta \phi=\phi-\phi^0$ and $\delta \cc = \cc - \cc^0$ obey the Laplace and diffusion equations respectively. Their boundary conditions read
\begin{eqnarray}
\partial_t \sigma =K \nabla_n \delta \phi +Kk_{11}\nabla^2_\perp \delta \phi+Kk_{12}\nabla^2_\perp \delta \cc
\label{dsigma}
\\
q\partial_t \gamma =K \nabla_n \delta \cc +Kk_{21}\nabla^2_\perp \delta \phi+Kk_{22}\nabla^2_\perp \delta \cc
\label{dgamma}
\end{eqnarray}

It is the first term $K \int_{\sp} \left( \delta \phi {\bf 1}-\br {\bf \nabla} \delta \phi \right) \cdot d {\bf A}$ in eq.(\ref{bdi}) which gives the main contribution in the salt polarization theory \cite{De Lacey}.

\subsection{Charging current density and the electric dipole}
It is instructive to examine the dipole produced by the charging current density  $\di$.
\begin{eqnarray}
\vtot \langle \delta i_a \rangle &=&\int_\vtot \delta i_a dV \nonumber \\
&=& \int_\vtot \nabla_b (r_a \delta i_b) dV-\int_\vtot r_a {\bf \nabla }\cdot \bi \ dV \nonumber \\
&=&-\int_\se r_a \delta \bi \cdot d \bA+\int_\vtot r_a \partial_t \rhof dV
\nonumber 
\end{eqnarray} 
where we have used the identity $i_a=\nabla_b(r_ai_b)-r_a \nabla_b i_b$ to get the second line. In the last step, Gauss's theorem and the charge conservation equation are used.
In the last term, we can restrict the range of integration to $\vin$ because the charge density $\rhof$ vanishes within $\vout$ for the symmetric electrolyte.
The result reads
\begin{eqnarray}
\partial_t {\bf P}_{tot} =\vtot \langle \delta \bi \rangle = -\int_\se \br \delta \bi \cdot d \bA + \int_\vin \br \partial_t \rhof dV
\label{p}
\end{eqnarray}
where ${\bf P}_{tot}$ is the dipole of the system. 
The first term $-\int_\se \br \delta \bi \cdot d \bA$ is the dipole increment due to the free charges flow into the electrode surface. 
The term $\int_\vin \br \partial_t \rho dV$ is the time derivative of the dipole close to or inside the particles. 
Compare eq.(\ref{bdi}) with eq.(\ref{p}), 
 we conclude that the large  polarization in the double layer polarization theory consists of the dipole built up by the (free) ions accumulated at the boundary of the electrodes, rather than the dipole around or within the particles.

\section{Static Dielectric Constant and the Free Energy Stored}

We now establish our main goal, the reversible work done at the charging process is due to the free energy stored as the form of the non-uniform salt concentration.  Our result also contains the additional contributions within the double layer regions.
The charging power is \cite{note}
\begin{eqnarray}
{\cal P}=  \int_\se \phi  \dei \cdot d \bA
\label{P}
\end{eqnarray}
Note that $\se$ has two components close to the two electrodes. Equation (\ref{P}) is identical to the charging current times the voltage difference between the two electrodes.
 
Because $\langle \bE \rangle=\langle \bE^0 \rangle $ and both $\phi$, $\phi^0$ are constant on each electrodes, the reference potential $\phi^0$ is the same as the true potential at the electrode surface $\se$. Therefore $\phi$ can be replaced by $\phi^0$  in eq.(\ref{P}).
In the bulk solution, the current is simply given as $\dei=-K{\bf \nabla }\delta \phi$.
Using that $\nabla^2 \phi^0=\nabla^2 \delta \phi=0$ within $\vout$, one integration by parts to get the identity
\begin{eqnarray}
 \int_\se \phi^0 \bnabla  \delta \phi \cdot d \bA+ \int_\sp \phi^0 \bnabla  \delta \phi \cdot d \bA=-\int_\vout \bnabla \cdot \left( \phi^0 \bnabla \delta \phi  \right) dV=-\int_\vout \bnabla \phi^0 \cdot \bnabla \delta \phi dV  
\nonumber \\ = - \int_\vout \bnabla \cdot \left( \delta \phi \bnabla \phi^0 \right) dV = \int_\sp \delta \phi \bnabla \phi^0 \cdot d \bA + \int_\se \delta \phi \bnabla \phi^0 \cdot d \bA
\nonumber
\end{eqnarray}
The last term is actually zero because $\delta \phi =0$ at the electrodes $\se$. Use the above identity, we obtain
\begin{eqnarray}
{\cal P}=-K \int_\se \phi^0 \bnabla \delta \phi \cdot  d \bA = K \int_\sp (\phi^0 {\bnabla }\delta \phi -\delta \phi {\bnabla} \phi^0) \cdot d \bA
\end{eqnarray}

We first rearrange the term $\int \phi^0 {\bf \nabla} \delta \phi \cdot d \bA$ so that the driving force of the surface current $\phi^0+(k_{12}/ k_{11}) \cc^0$ appears (see the last two terms of eq.(\ref{sigma0})).
\begin{eqnarray}
& &\int_\sp \phi^0 {\bnabla} \delta \phi \cdot d \bA
=\int_\sp \left( \phi^0 +\frac{k_{12}}{k_{11}}\cc^0 \right) {\bnabla} \delta \phi \cdot d \bA
-\frac{k_{12}}{k_{11}}\int_\sp \cc^0  {\bnabla} \delta \phi \cdot d \bA
\nonumber
\end{eqnarray}
In the first term, $\bnabla \delta \phi \cdot d \bA= \nabla_n \delta \phi dA$, and we eliminate $\nabla_n \delta \phi$ by eq.(\ref{dsigma}), to get
\begin{eqnarray}
& &\frac{1}{K}\int_\sp \left( \phi^0 +\frac{k_{12}}{k_{11}}\cc^0 \right) \partial_t \sigma dA
-k_{11} \int_\sp \left( \phi^0 +\frac{k_{12}}{k_{11}}\cc^0 \right)
\nabla^2_\perp \left( \delta \phi +\frac{k_{12}}{k_{11}} \delta \cc \right) dA
-\frac{k_{12}}{k_{11}}\int_\sp \cc^0  {\bf \nabla} \delta \phi \cdot d \bA
\nonumber \\
& & =\frac{1}{K}\int_\sp \left( \phi^0 +\frac{k_{12}}{k_{11}}\cc^0 \right) \partial_t \sigma dA
-k_{11} \int_\sp \left(\delta \phi +\frac{k_{12}}{k_{11}}\delta \cc \right)
\nabla^2_\perp \left(  \phi^0 +\frac{k_{12}}{k_{11}}  \cc^0 \right) dA
-\frac{k_{12}}{k_{11}}\int_\sp \cc^0  {\bnabla} \delta \phi \cdot d \bA
\nonumber 
\end{eqnarray}
where  (surface) integration by parts is used twice in the second term. (Note that $\sp$ is a close surface, and has no boundary.) Finally we use eq.(\ref{sigma0}) to eliminate $\phi^0+(k_{12}/k_{11}) \cc^0$ in the second term to get 
\begin{eqnarray}
\frac{1}{K}\int_\sp \left( \phi^0 +\frac{k_{12}}{k_{11}}\cc^0 \right) \partial_t \sigma dA
+ \int_\sp \left(\delta \phi +\frac{k_{12}}{k_{11}}\delta \cc \right)
\nabla_n \phi^0 dA
-\frac{k_{12}}{k_{11}}\int_\sp \cc^0  { \nabla_n} \delta \phi  dA
\nonumber 
\end{eqnarray}
 The above manipulations lead to
\begin{eqnarray}
{\cal P}=K\frac{k_{12}}{k_{11}}  
\int_\sp \left( \delta \cc \nabla_n \phi^0 -\cc^0 \nabla_n \delta \phi +K^{-1}\cc^0 \partial_t \sigma
\right) dA
 +\int_\sp \phi^0 \partial_t \sigma dA
\nonumber
\end{eqnarray}

The first term can be further transformed by using eqs.(\ref{sigma0})(\ref{dsigma}) to eliminate $\nabla_n \phi^0$, $\partial_t \sigma -K \nabla_n \delta \phi$ respectively, as
\begin{eqnarray}  
& &\int_\sp \left( \delta \cc \nabla_n \phi^0+\cc^0(K^{-1} \partial_t \sigma-\nabla_n \delta \phi)
\right) dA \nonumber \\
&=&k_{11}\int_\sp(- \delta \cc \nabla^2_\perp \phi^0 +\cc^0 \nabla^2_\perp \delta \phi ) dA \nonumber \\
&=&\frac{k_{11}}{k_{21}} \int_\sp (\delta \cc \nabla_n \cc^0-\cc^0 \nabla_n \delta \cc +K^{-1}\cc^0 q \partial_t \gamma) dA 
\nonumber 
\end{eqnarray}
where $\int_\sp \delta \cc \nabla^2_\perp \cc^0 dA =\int_\sp \cc^0 \nabla^2_\perp \delta \cc dA$ is used to cancel terms.
In the last step, (\ref{gamma0})(\ref{dgamma}) are used to eliminate $\nabla^2_\perp \phi^0$, $\nabla^2_\perp \delta \phi$ respectively.

We would like to enlarge the integration area of the first two terms to include $\se$, so that the surface integral on $\se$ and $\sp$ can be transformed to the bulk integral on $\vout $ by the Green's identity.
The salt concentration $\cc^0$ contributed from each particle decays as $1/r^2$ or faster, because the monopole strength vanishes  $\oint d \bA \cdot \bnabla \cc^0 =0$ (we use eq.(\ref{gamma0}), and integration by parts). The quantity $\delta \cc$ decays no slower than $\cc^0$.  Therefore the integral on the electrode surface $\se$ scales less than the system size. Whereas the original integrals on $\sp$ are extensive, and hence dominant. For a large enough  system (thermodynamic limit), we freely enlarge the integration area of the first two terms as $\sp+\se$ with negligible error, to derive
\begin{eqnarray}
& &\int_{\sp+\se} (\delta \cc \nabla_n \cc^0-\cc^0 \nabla_n \delta \cc ) dA 
\nonumber \\
&=&\int_\vout (-\delta \cc \nabla^2 \cc^0 + \cc^0 \nabla^2 \delta \cc)dV \nonumber \\
&=&D^{-1}\int_\vout \cc^0 \partial_t \delta \cc dV 
\nonumber
\end{eqnarray}
where we have used the Green's identity,  $\nabla^2\cc^0=0$, and $\partial_t \delta \cc=D\nabla^2\delta \cc$.

Finally, put the above derivations together, and use the Onsager relation $k_{12}=k_{21}$, we obtain the result 
\begin{eqnarray}
&{\cal P}&=V_{tot} \E \cdot \di \nonumber \\
&=&\frac{K}{D} \int_\vout \cc^0 \partial_t \delta \cc dV
+ \int_\sp \cc^0 q \partial_t \gamma dA + \int_\sp \phi^0 \partial_t \sigma dA
\nonumber \\
&=&2 \int_\vout \frac{kT \delta c}{ C} \partial_t \delta c +
\int_\sp \frac{kT \delta c}{ C} \partial_t \gamma dA 
+ \int_\sp \phi \partial_t \sigma dA
\label{ptot}
\end{eqnarray}
where in the last step, we have assumed that the charging rate is very small, to neglect the terms quadratic in the charging rate.

In the linear response regime, the perturbations $\gamma$, $\sigma$, $\delta c$, and $\phi$ are all proportional. We can integrate eq.(\ref{ptot}) to obtain
\begin{eqnarray}
\vtot \frac{\Delta \epsilon }{2} E^2= \Delta F
=\int_\vout  \frac{kT(\delta c)^2}{ C}  dV+\int_\sp \frac{kT \delta c  }{2  C} \gamma dA 
+ \int_\sp \frac{\phi  }{2} \sigma  dA
\label{e0}
\end{eqnarray}
Equation (\ref{e0}) shows that the external electric work done of the charging process is actually saved in the forms of the non-uniform salt concentration in the bulk (first term) and inside the double layer (second term), as well as the electric energy stored in the polarized double layer (last term). 
The first term is as proposed by Grosse and Shilov \cite{Grosse97}. The second term is of the similar physical origin as the first term but its magnitude is much smaller for large particles with thin double layers. The dynamic calculations \cite{Dukhin,Fixman} do not have this term because they neglect the (small) term $\partial_t \gamma$ in eq.(\ref{gamma}) so that their static result is less complete as in eq.(\ref{e0}). The last term accounts for the energy stored in the interface polarization \cite{O'Brien86}. It is not included in the dynamic calculation \cite{Dukhin,Fixman} because the term $\partial_t \sigma$ in eq.(\ref{sigma}) has been neglected.   The exact values of $\sigma$ and $\gamma$ depend on the quantities like the dielectric constant and the conductivity of the particles, but the general forms of the last two terms in eq.(\ref{e0}) remain the same.

\section{Summary}

Using the electrokinetic equations, we have shown in eq.(\ref{e0}) that the work done on the charging current is stored in (i) the  free energy of the uneven salt concentration in the bulk solution, (ii) the free energy of the double layer excess ion density, and (iii) the polarized double layer charge.

We also show in eqs.(\ref{bdi})(\ref{p}) that,  the dipole of the free charge close to the electrode surface, gives the dominant dielectric increment of the double layer polarization theory.

{\bf Acknowledgments}

This work is funded by the National Science Council of Taiwan.

\pagebreak

\section*{ Figure caption}

{\bf Figure 1.} Two surfaces $\sp$ and $\se$ are shown. The space $\vout$ is bounded between the two surfaces. The space $\vin$ is within the surface $\sp$, which includes the charged particles and their double layers.

\pagebreak
{\begin{figure}
\begin{center}
\epsfig{file=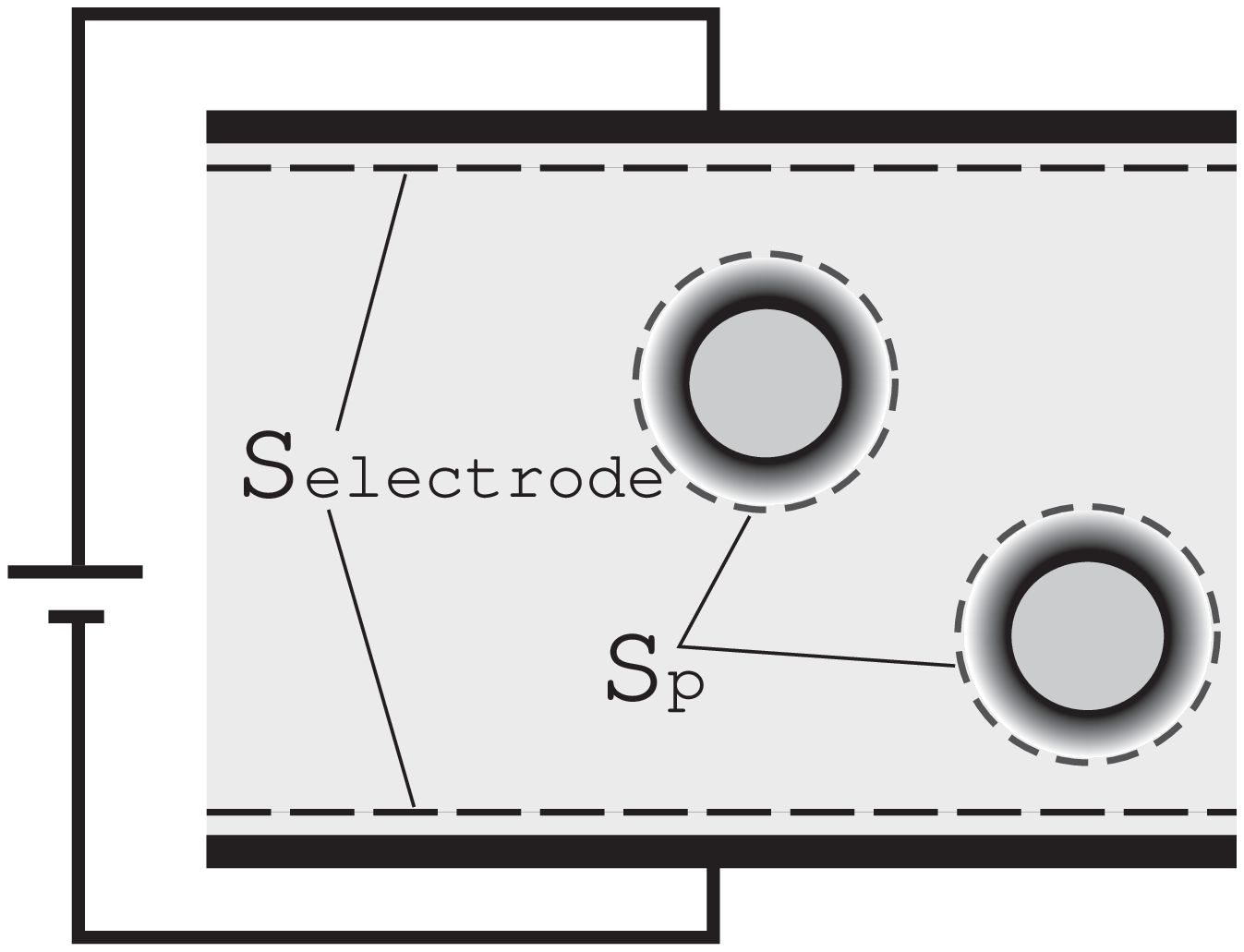, height=10cm}
\end{center}
\end{figure}}



\begin{thebibliography}{9}
\bibitem{note} Because $\phi$ is constant on the flat electrodes, $\int_\se \phi \delta \bi \cdot d \bA =\di \cdot \int_\se \phi d \bA = -\di \cdot  \int_\vtot {\bnabla } \phi dV= \vtot \E \cdot \di$.
\end{thebibliography}
\end{document}